\documentclass[preprint,preprintnumbers,amsmath,amssymb]{revtex4-1}
\usepackage{txfonts}
\usepackage{amsfonts}
\usepackage{amsmath}
\usepackage{graphicx} 
\usepackage{dcolumn}  
\usepackage{bm}
\usepackage{overpic}
\usepackage{color}
\begin{document}

\title{Frequency-dependent attenuation and elasticity in
  unconsolidated earth materials: effect of damping}

\author{Yanqing Hu$^1$, Hern\'an A. Makse$^{1,3}$, John
  J. Valenza$^2$, David L. Johnson$^2$}

\affiliation{$^1$ Levich Institute and Physics Department, City
  College of New York, New York, New York 10031, USA\\ $^2$
  Schlumberger-Doll Research, One Hampshire, Cambridge, Massachusetts
  02139, USA\\ $^3$ Corresponding author}

\begin{abstract}

{\bf We use the Discrete Element Method (DEM) to understand the
  underlying attenuation mechanism in granular media, with special
  applicability to the measurements of the so-called effective mass
  developed earlier.  We consider that the particles interact via
  Hertz-Mindlin elastic contact forces and that the damping is
  describable as a force proportional to the velocity difference of
  contacting grains.  We determine the behavior of the complex-valued
  normal mode frequencies using 1) DEM, 2) direct diagonalization of
  the relevant matrix, and 3) a numerical search for the zeros of the
  relevant determinant.  All three methods are in strong agreement
  with each other.  The real and the imaginary parts of each normal
  mode frequency characterize the elastic and the dissipative
  properties, respectively, of the granular medium.  We demonstrate
  that, as the interparticle damping, $\xi$, increases, the normal
  modes exhibit nearly circular trajectories in the complex frequency
  plane and that for a given value of $\xi$ they all lie on or near a
  circle of radius $R$ centered on the point $-iR$ in the complex
  plane, where $R\propto 1/\xi$.  We show that each normal mode
  becomes critically damped at a value of the damping parameter $\xi
  \approx 1/\omega_n^0$, where $\omega_n^0$ is the (real-valued)
  frequency when there is no damping.  The strong indication is that
  these conclusions carry over to the properties of real granular
  media whose dissipation is dominated by the relative motion of
  contacting grains.   For example, compressional or shear waves in
  unconsolidated dry sediments can be expected to become overdamped
  beyond a critical frequency, depending upon the strength of the
  intergranular damping constant.}

\end{abstract}
\maketitle

\begin{center}
\begin{large}
{\bf INTRODUCTION}
\end{large}
\end{center}

An important challenge in the development of a hydrocarbon reservoir
is optimizing well completion and production.  After placing a well,
sonic well logging techniques characterize the subsurface at high
resolution in the immediate vicinity of a borehole.
For a given strata, the acoustic properties depend on many
characteristics (composition, structure, porosity) and conditions
(stress, temperature, pore fluid).  In the soft unconsolidated
formations of interest in the present study, sonic measurements are
further complicated by nonlinear effects arising from heterogeneities,
inhomogeneous stress distributions, and dissipation due to fluid flow
(Domenico, 1977; O`Connell and Budiansky, 1977; Digby, 1981; Winkler,
1983; Walton, 1987; Chen et al., 1988; Goddard, 1990; de Gennes, 1996;
Behringer an Jenkins, 1997; Norris and Johnson, 1997; Johnson et al.,
1998; Guyer and Johnson, 1999; Makse et al., 1999). These
nonlinearities have important effects on the dynamics of
unconsolidated media as reviewed by Guyer and Johnson (1999) and
Johnson et al. (1996).  Therefore, coupled with a representative
model of the granular medium (Goddard, 1990; Guyer and Johnson, 1999;
Domenico, 1977; Winkler, 1983; Behringer and Jenkins, 1997; Norris and
Johnson, 1997) sonic wave propagation can be utilized to characterize
subsurface conditions and elucidate active mechanisms. In turn, this
information can be utilized during well completion and production.
Another important effect of non-linearities is to cause dispersion in
the system as studied in Johnson et al. (1996) and Guyer and
Johnson (1999).

The motivation of the present study is to develop theoretical and
numerical tools to study the acoustic and dissipative properties
of unconsolidated granular materials. Our findings can then be employed
to a) improve the interpretation of sonic logging measurements, and
(b) to optimize granular media for the dissipation of acoustic energy.
First, we utilize the Discrete Element Methods (DEM) to test the
theory for the effective mass of discrete systems as outlined in
Valenza and Johnson (2012).  We utilize this DEM and analytical
framework to calculate the frequency-dependent effective mass of
unconsolidated granular media held in a cup, and study the effect of
interparticle damping on the normal modes in the system.


Previous work on the normal modes of unconsolidated granular materials (Alexander, 1998; O`Hern et al., 2003; Silbert at al., 2005; Somfai et al., 2007; Wyart et al., 2005a,b)  have studied the vibrational density of states of a granular system as
the external applied pressure is diminished and the volume fraction of
the system decreases towards the point of random close packing (RCP),
i.e., when the granular medium is a fragile unconsolidated formation  (Makse et al., 2000) and (O`Hern et al., 2003).  However, these previous studies have not
considered the important effect of attenuation on the normal
modes. Attenuation plays a crucial role in governing the dynamic
response of real system.
Thus, interpretation of any experiment and acoustic logging needs to
take into account the dissipative nature of granular
matter. 

In a previous study we have focused on the stress-dependent
dissipative characteristics of the granular medium (Hu et al., 2014).  The
effect of attenuation at the grain-grain contacts has been also
studied experimentally and theoretically in Valenza and Johnson (2012).  Here, we test those theoretical developments with Discrete Element Methods (DEM) (Cundall, 1979) with a particular focus on the low-frequency
      modes in the effective mass of granular materials.
      We further develop this theoretical basis for granular
      systems in order to investigate the dependence of the normal
      modes on the damping mechanism at the interparticle contact. An
      area of primary focus is the consequences of assuming the
      elastic and damping matrices commute.

The paper is organized as follows: First we introduce
the concept of the effective mass and then we review specific relevant theoretical results, in particular the relation
between the effective mass and the normal modes of vibration in the system.  In the special case that the damping matrix and the stiffness matrix are proportional to each other with proportionality $\xi$, we derive properties of the trajectories of the normal mode frequencies, in the complex plane, as a function of $\xi$.  Next, we outline the DEM computational technique as well as the eigenvalue-eigenvector technique with which we compute the main results of this paper.  Those results are analyzed in detail in view of the previously mentioned exact and approximate results and we conclude with our summarizing remarks.

\begin{center}
\begin{large}
{\bf EFFECTIVE MASS OF A GRANULAR MEDIUM}
\end{large}
\end{center}

\begin{figure}
\centering{
  \includegraphics[width=0.9\columnwidth]{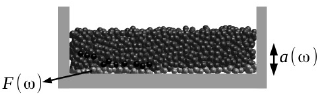}}
\caption{}
\label{fig:packing}
\end{figure}

The effective mass of the granular medium in a cup of mass $M_c$
vibrating with frequency $\omega$ is defined as (Hsu et al., 2009):

\begin{equation}
  \tilde{M}(\omega)
= \frac{F(\omega)}{a(\omega)} - M_c.
\label{meff}
\end{equation}
For each frequency $\omega$, $F(\omega)$ is determined by measuring
the force necessary to displace the cup and, $a(\omega)$ is the
associated acceleration. The system is schematically explained in
Fig. \ref{fig:packing}. Experimentally, both quantities are recorded
in the frequency domain after waiting a short stabilization period of
a few seconds. Then, equation \ref{meff} is employed to construct
$\tilde{M}(\omega)$.  In general, $\tilde{M}(\omega) = M_1(\omega)
\:+\: iM_2(\omega)$ is complex-valued and reflects the partially
in-phase and out-of-phase motion of the individual grains relative to
the cup motion. $M_1$ and $M_2$ characterize the ensemble averaged
elastic and dissipative capacity of the material in a simplified
continuum with negligible boundary effects (see for instance Henann et al. (2013).   An ideal example is provided by a low viscosity liquid
where the dimensions of the cup are much larger than the viscous skin
depth across the experimental frequency band; In this case the
effective mass resonates at odd-multiples of $1/4$ wavelength of the
longitudinal wave in the medium (Hsu et al., 2009).  In fact, the
dominant low frequency normal modes in a granular medium exhibit
similar scaling behavior. However, the scaling of the modes in a
confined granular medium is effected by the preparation protocol, and
boundary condition  (Henann et al., 2013).

\begin{figure}
\centering{\includegraphics[width=0.9\columnwidth]{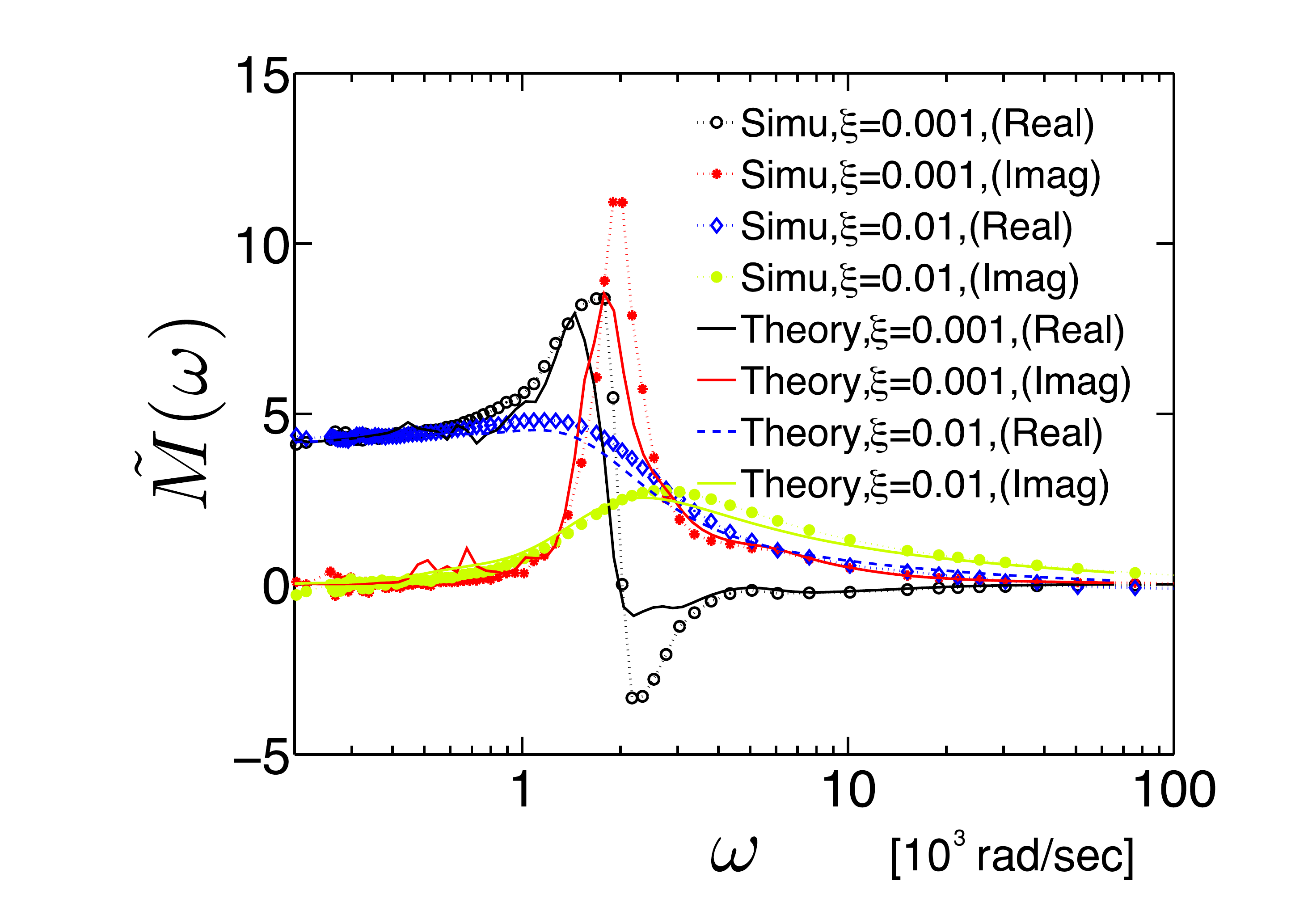}}
\caption{}
\label{TheorySim}
\end{figure}

The imaginary part of the frequency dependent effective mass in a
granular medium is roughly proportional to the frequency dependent
attenuation in the system. This dissipation can be due to solid
contact friction (asperities, material plasticity, etc.) or wetting
dynamics from liquid bridges or films between grains. Figure
\ref{TheorySim} shows the results of a frequency sweep of the real
($M_1$) and imaginary ($M_2$) components of the effective mass
obtained with DEM for packings under gravity with the indicated
damping coefficients (DEM procedure outlined below).  These results
are qualitatively similar to those determined experimentally, and
reported in Hsu et al. (2009) and Valenza and Johnson (2012).  Here, we adapt
the DEM technique to obtain the low-frequency modes in the region of
interest. In a manner analogous to that achieved experimentally in
Valenza and Johnson (2012) we use DEM to test the effect of interparticle
damping on the dynamics of discrete systems.


\begin{center}
\begin{large}
{\bf PREVIOUS THEORETICAL RESULTS}
\end{large}
\end{center}

{\bf Hertz-Mindlin interparticle force law}

We consider a granular medium made of spherical particles interacting
via Hertz-Mindlin contact forces and dissipative
forces at the contact point.  Our technique is generalizable, however, to nonspherical particles  (Baule, et al., 2013; Baule and Makse,
2014).  The normal component of the contact
force between any two contacting particles with radius $R$ is the
Hertz force (Landau and Lifschitz, 1970; Johnson, 1985) derived as:

\begin{equation}
F_n=\frac{2}{3}k_{n}R^{1/2}x_{ij}^{3/2},
\label{n}
\end{equation}

where the normal deformation (1/2 the overlap between the spheres)
between the neighboring grains is $x_{ij}=\frac{1}{2}[2
  R-|\mathbf{x}_i-\mathbf{x}_j|]$, $\mathbf{x}_{i,j}$ are the position
vectors, and $k_n$ is the normal spring constant. The latter is
defined in terms of the corresponding material properties. The normal
elastic constant is $k_n= 4 G_g / (1-\nu_g),$ where $G_g$ is the shear
modulus, and $\nu_g$ is the Poisson's ratio of the material from which
the grains are made.

The tangential Mindlin force between neighboring grains in contact is (Shafer et al., 1996):

 \begin{equation}
\Delta F_t = k_t (R x_{ij})^{1/2} \Delta s,
\label{t}
\end{equation}
where $k_t = 8 G_g / (2-\nu_g),$ is the tangential spring constant,
and the variable $s$ is defined such that the relative shear
displacement between the two grain centers is $2s$.

In addition to these elastic forces we consider the effects of
intergranular damping which are manifest here as forces proportional
to the difference between the velocity values of two contacting
grains.  The general linearized form is $F_{dis}^N \propto
\dot{x}_{ij} $ and $F_{dis}^T \propto \dot{s}_{ij} $.

Finally, Coulomb friction with interparticle friction coefficient
$\mu$ imposes $F_t \le \mu F_n,$ at every contact.  The experimental
signature of this effect is that if the amplitude of vibration is
large enough that adjacent grains slide against each other the
measured effective mass becomes amplitude dependent.  The previously
cited experiments carefully avoided this situation by sticking to low
amplitude vibrations.  Accordingly, in the present article we shall
assume either that there is no relative tangential slip between grains
(perfect stick) or that there is no tangential force at all (perfect
slip), as the case may be.


We are interested in the linear response of the granular system to
infinitesimal perturbations. Therefore we consider the linearized
forms of the above equations, ie, we use the Hertz-Mindlin force law
linearized about the static value of the normal compression of
contacts.  The resulting elastic stiffness is the slope of the force
law evaluated at $x_{ij}$ (eg. equation \ref{n} for the normal component) and equation \ref{t} for the transverse component:
 \begin{equation} \label{Hertz}
k^N(x_{ij}) =k_n R^{1/2} x_{ij}^{1/2},
\end{equation}
and
 \begin{equation}
k^T(x_{ij}) =k_t R^{1/2} x_{ij}^{1/2}.
\label{Mindlin}
\end{equation}

Our goal here is to understand the effects of dissipation when, for
instance the particles are coated with a viscous fluid, such as
described in the experiments of Valenza and Johnson (2012).  The
dissipation is mainly due to the 'squirt flow' in liquid bridges at
the particle contacts.  We shall assume for simplicity that the
damping constants are independent of the particle deformation and are
therefore the same for all particles that are in contact. It shall
prove to be convenient to write the damping forces as
\begin{equation}
F_{diss}^N = \xi \,\, \overline{k^N} \,\, \dot{x}_{ij},
\label{disn}
\end{equation}
and
\begin{equation}
F_{diss}^T = \xi \,\, \overline{k^T}  \,\, \dot{s}.
\label{dist}
\end{equation}
where the overbar signifies an average over all non-zero contacts and
$\xi$, which has the dimensions of time, will be our control
parameter.  It is a stand-in for the viscosity of the fluid at the
grain-grain contacts.



{\bf Theory of normal modes}

In a typical experiment or simulation, the grains are settled in a cup
that moves sinusoidally with magnitude $W$, taken in the
$z-$direction.  We denote $\mathbf{W}=W \mathbf{\hat {z}}$ as the
uniaxial displacement of the cup. In the
analogous physical experiments (Hsu et al., 2009) $W$ is less than $1 \mu$m,
at least three orders of magnitude smaller than $R$. This assures that
the external displacement is smaller than the static compression at
the grain contacts.  Therefore, we take all $\mathbf{u}_i$ to be infinitesimal,
and utilize the linear equation of motion for the $i$-th particle with mass
$m$ to describe the system dynamics.  The
coupled equations of motion for the degrees of freedom in a system of
$N$ particles is written as [see Valenza and Johnson (2012) and Hu et al. (2014) for details]:
 \begin{equation} \label{h}
H_{ij}(\omega) u_j = K_{i\omega} W, \,\,\,\,\,\, \mbox{$i,j = 1 :
  6N$},
\end{equation}
where $i$ and $j$ label each of the $6N$ degrees of freedom in the problem. The vector $\{u_j\}$ accounts for the set of $3N$ particle displacements and $3N$ particle rotations.   The dynamical matrix $\mathbf{H}_{ij}(\omega)$ in the frequency domain
can be written as,
 \begin{equation}
H_{ij}(\omega)=- m_{ij} \omega^2 - i \omega B_{ij} + K_{ij}.
\label{matrix}
\end{equation}
Here the mass matrix $\mathbf{M}$ has elements $m_{ij} = m
\delta_{ij}$, with $\mathbf{I}$ the identity matrix if the particles
interact via central forces only (frictionless), and it contains the
moment of inertia for rotating particles interacting via tangential
forces. Each term in $H_{ij}$ accounts for the inertial term, the
dissipative matrix $\mathbf{B}$, and elastic matrix $\mathbf{K}$,
defined at the contact between particles $p$ and $q$,
respectively. The elements of the elastic matrix $K_{ij}$ follow from the Hertz-Mindlin stiffness constants for infinitesimal displacement, equations \ref{Hertz} and \ref{Mindlin}.  It is convenient to write this matrix for a given pair of grains $p$ and $q$ that are in contact as
 \begin{equation}
\mathbf{K}_{pq}= k^N(x_{pq})\mathbf{\hat{d}}_{pq}\mathbf{\hat{d}}_{pq}
+ k^T(x_{pq}) [\mathbf{I} -
  \mathbf{\hat{d}}_{pq}\mathbf{\hat{d}}_{pq}],
\label{elastic}
\end{equation}
where $\mathbf{\hat{d}}_{pq}$ denotes the direction of the normal
displacement along the contact point and we are using dyadic notation.
Similarly, the damping matrix $B_{pq}$ follows from equations \ref{disn} and \ref{dist} and may be written as
 \begin{equation} \label{Ndamp1}
\mathbf{B}_{pq}=\xi\; \overline{k^N}\;\mathbf{\hat{d}}_{pq}\mathbf{\hat{d}}_{pq}
+ \xi\; \overline{k^T}\;[\mathbf{I} -
  \mathbf{\hat{d}}_{pq}\mathbf{\hat{d}}_{pq}]\:\:\:.
\end{equation}
If the particles, $p$ and $q$, are not in contact
($|\mathbf{x}_p-\mathbf{x}_q|>2R$), we set both $\mathbf{K}_{pq}=0$
and $\mathbf{B}_{pq}=0$.
In equation \ref{h} $K_{iw}$ is the
generalized spring constant connecting a particle to the walls of the
cup which moves oscillatory in the $z$ direction with amplitude $W$.

Inverting the matrix $\mathbf{H}$ we obtain the effective mass as has
been shown in our previous work (Valenza and Johnson, 2012):
 \begin{equation} \label{effmass}
\tilde{M}(\omega)=m_i
    [H^{-1}(\omega)]_{ij} K_{jw}.
\end{equation}
%
This equation expresses the relation between the effective mass and
the normal mode spectrum. The peaks observed in the effective mass
(Fig. \ref{TheorySim}), $\tilde{M}(\omega)$, are due to the set of
normal modes, $e_j^n$, that are a solution to equation \ref{h} when there
is no forcing by the cup, $W=0$, i.e.,
 \begin{equation}
H_{ij}(\omega_n) e_j^n = 0.
\label{eigen}
\end{equation}
Thus, the normal modes, $ e_j^n$, are those eigenvectors of
$\mathbf{H}$ for which the corresponding eigenvalue is zero, and they
occur at specific complex-valued frequencies, $\omega_n$.

\begin{center}
\begin{large}
{\bf EIGENVALUE TRAJECTORIES AS A FUNCTION OF DAMPING}
\end{large}
\end{center}

While previous theoretical studies have considered the undamped modes
of frictionless systems (Wyart et al., 2005a,b), the
interpretation of the effective mass data from experiments or sonic
logging data necessitates a formalism that considers damping and
rotational modes that are indigenous in to real granular matter. Our
formalism thus generalizes previous results valid for frictionless
undamped systems to more realistic granular materials consisting of
dissipative interactions with tangential forces.

Equation \ref{effmass} implies that $\omega_n$ is the resonance
frequency when $\mathbf{H}(\omega)$ has at least one zero eigenvalue
or the determinant of $\mathbf{H}(\omega)$ is equal to zero. When
$\xi>0$, the resonance frequency is a complex number and the imaginary
part is relevant to the attenuative properties of the granular
system. Here, we study the effect of interparticle damping, $\xi$, on
the resonant modes $\omega_n$.

We are interested in the complete set $\{e_j^n\}$ of complex valued
normal modes that satisfy equation \ref{eigen}.  For these modes, using some results from Caughey (1960) and Hu et al. (2014) we can
rewrite equation \ref{eigen} in the frequency domain as:
 \begin{equation}
\left( - \mathbf{I} \omega_{n}^{2} - i\omega_{n} \tilde{\mathbf{B}} +
\tilde{\mathbf{K}} \right) \mathbf{q} = 0.
\label{19}
\end{equation}
Here we have defined: $\mathbf{q} \equiv \mathbf{M}^{-1/2}
\mathbf{e}$,
$\tilde{\mathbf{B}}\equiv\mathbf{M}^{-1/2}\mathbf{B}\mathbf{M}^{-1/2}$
and
$\tilde{\mathbf{K}}\equiv\mathbf{M}^{-1/2}\mathbf{K}\mathbf{M}^{-1/2}$,
provided that the $\mathbf{M}$ matrix is positive definite, we can
always find $\mathbf{M}^{-1/2}$.




It is obvious from their definitions,
  equations \ref{elastic} and \ref{Ndamp1}, that the matrices {\bf K}
  and {\bf B} do not commute.  It shall prove to be informative to
  explore the consequences of assuming not only that they commute but
  that they are proportional to each other, viz:

 \begin{equation}
\mathbf{B}_{ij} \simeq \xi \,\, \mathbf{K}_{ij}.
\label{propto}
\end{equation}

The distribution of normal stiffness $k^N(x_{ij})$ is directly related
to the distribution of interparticle normal forces via the
distribution of deformation $x_{ij}$. Such a force distribution is
well known to display an exponential tail and to be relatively homogenous (Makse et al., 2000).  Therefore, equation \ref{propto} amounts to neglecting the
inhomogeneities in the force distribution as an approximation for the
elastic and damping matrices. If each contact stiffness value was replaced by the average thereof, equation \ref{propto} would be exact.



An important implication of equation \ref{propto} is that the eigenvectors of  $\tilde{\bf K}$ and $\tilde{\bf B}$  are the same.  The
normal modes in the damped system are exactly the same as in the
undamped case except that they now have complex-valued frequencies,
due to the attenuation.
Here, we study the effect of the commuting approximation on the behavior of the
normal mode frequencies as $\xi$ is varied.

Let $\omega_{n0}^{2}>0$ be the eigenvalues of $\tilde{\mathbf{K}}$
with $\mathbf{q^n}$ the corresponding eigenvectors. Thus, in the
absence of attenuation $\tilde{\mathbf{B}}=0$, $\pm \omega_{n0}$ is
the undamped frequency of oscillation of this mode. If
$\tilde{\mathbf{B}}$ and $\tilde{\mathbf{K}}$ are proportional, then
each mode exactly decouples and we can re-write equation \ref{19} as:
 \begin{equation}
\Big[-\omega_{n}^{2} - i \omega_n \xi \omega_{n0}^{2} + \omega_{n0}^2
  \Big] \mathbf{q^n} = 0,
\label{23}
\end{equation}
which is a quadratic equation in $\omega_n$ with roots:
 \begin{equation}
\omega_n = -i \frac{\xi \omega_{n0}^{2}}{2} \pm \omega_{n0}\sqrt{1 -
  \left( \frac{\xi \omega_{n0}}{2}\right)^2}.
\label{eqn:roots_omegan}
\end{equation}
From this result it is clear that, as the damping parameter, $\xi$, is
varied, each complex-valued normal mode frequency, $\omega_n(\xi)$,
follows a trajectory which is a circle of radius $\omega_{n0}$ as long
as the damping is smaller than a critical value given by

%
 \begin{equation}
\xi_c = \frac{2}{\omega_{n0}}.
\label{eq:scaling_xi_c}
\end{equation}
For large enough damping $\xi > \xi_c$, each normal mode becomes
overdamped, and the corresponding frequencies are purely imaginary
valued (Inman and Andry, 1980; Bhaskar, 1997).  For $\xi < \xi_c$, the modal
frequencies are damped oscillators with
 \begin{equation}
|\omega_n(\xi)| = \omega_{n0}.
\label{circular}
\end{equation}
That is, the trajectories of the modes in the plane
$\Big($Re$[\omega_n(\xi)],$ Im$[\omega_n(\xi)]\Big)$ are exactly
circular as a function of $\xi$.

For a fixed value of $\xi$ we note the following consequence of equation \ref{eqn:roots_omegan} as long as $\xi<\xi_c$:
 \begin{equation}
{\rm Re}[\omega_n]^2 + \Big({\rm Im}[\omega_n] + \frac{1}{\xi}\Big)^2
= \frac{1}{\xi^2}.
\label{circle}
\end{equation}
That is, a cross-plot of the real and the imaginary parts of all the normal modes forms a circle centered at
$\Big($Re$[\omega_n],$ Im$[\omega_n]\Big) = (0, -1/\xi)$ having radius $1/\xi$.

If we take the low dissipation limit of equation \ref{eqn:roots_omegan},
 i.e. $\xi \rightarrow 0$ and expand the square root term, we see that
 the normal modes correspond to the purely elastic system
 $\omega_{n0}$ plus an imaginary part, proportional to the damping
 parameter:
 \begin{equation}
\omega_n \simeq \omega_{n0} - i\frac{\xi \omega_{n0}^2}{2} + O(\xi^2).
\end{equation}
In the same limit, we observe that the real and imaginary parts of the
normal modes follow a parabola in the complex plane for a given small
fixed $\xi$:
 \begin{equation}
{\rm Im}[\omega_n] \simeq - \frac{\xi}{2} {\rm Re}[\omega_n]^2 + O(\xi^2).
\label{parabola}
\end{equation}



Equation \ref{eq:scaling_xi_c} can be written in adimensional form.
We normalize the variables by $\omega_{n0}$, such that $\omega'_{n} =
\omega_n / \omega_{n0}$ and $\xi'=\xi \omega_{n0}$, then
equation \ref{eqn:roots_omegan} becomes adimensional:
$$
\omega'_n = -i \frac{\xi'}{2} \pm \sqrt{1 -
  \left( \frac{\xi'}{2}\right)^2}.
$$ In this case, equation \ref{eq:scaling_xi_c} for the critical damping
parameter becomes: $\xi'_c=2.$


We emphasize that equations \ref{23} - \ref{parabola} all follow
directly from the assumption of the validity of equation \ref{propto}.
Inasmuch as our {\it ansatz}, equations \ref{elastic} and \ref{Ndamp1},
does not strictly obey equation \ref{propto} it shall prove instructive
to examine how well our results approximately obey equations \ref{23} -
\ref{parabola}.

\begin{center}
\begin{large}
{\bf DEM SIMULATIONS}
\end{large}
\end{center}

The primary motivation of this work is to test the predictions of
equations \ref{eq:scaling_xi_c}, \ref{circle} and \ref{parabola}
using the DEM.  The discrete simulations consist of a monodisperse
granular packing where the grains interact via the contact force laws
defined in the previous sections.  For all calculations the particles
are homogeneous with radius $R=1$mm, particle density $\rho=10^3$ kg
m$^{-3}$, shear modulus $G_g=161$ GPa and Poisson ratio $v_g=0.2$. We
prepare a granular packing and impose gravity $g=9.8$m/s$^2$ in the
direction perpendicular to the bottom wall.

The preparation protocols used to generate packings are similar to those used in
Makse et al. (1999); Makse et al. (2000); Makse et al. (2004); Bruji\'c
et al. (2005); Bruji\'c et al. (2007); Magnanimo, et al. (2008); Song et
al. (2008); Jin and Makse (2010) where it is demonstrated that our
approach generates stable jammed packings.  Here, we generate a 3D distribution of spheres  using a random number generator to locate the sphere centers.  If an added sphere overlaps one or more existing ones, it is discarded from the ensemble.  Periodic boundary conditions are assumed in all 3 dimensions at this step, using a periodicity which is large as compared to a sphere diameter.  Next, the dimensions of the periodic unit cell are smoothly and uniformly decreased causing the spheres to rearrange under the influence of the  intergranular forces.  At this stage we assume the spheres interact by Hertz central forces only; this assumption mimics the effects of a vibratory filling protocol in a real experiment and it guarantees that the spheres form a random close packed structure.   Once a stable configuration corresponding to a finite, but small, confining pressure is established we consider that all the spheres which intersect two opposing  walls of the unit cell are now taken  to be permanently fixed, and these spheres become the ``top" and `bottom" of the ensemble.  The force of gravity is now introduced pointing  in the bottom direction and the top wall is removed.  Those spheres which do not comprise the bottom are allowed to rearrange again and this becomes the starting configuration for our dynamical calculations.  We still maintain periodic
boundary conditions in both directions perpendicular to
gravity.  Figure ~\ref{fig:packing} shows a
typical configuration of the packing.

Damping at the grain-grain contact is governed by
 equation \ref{Ndamp1} with $\xi$ as the damping
parameter.  We measure the damping parameter in ms and the frequencies
in $10^3$rad/sec.
The packings are composed of $N$ particles and we vary $N$ from small
systems of 14 particles up to $N=400$ to test different aspects of the
theory. The small system of 14 particles is used only to test the
calculation of the normal modes.  

First we validate our DEM framework by comparing the dynamical
effective mass to the analytical prediction of
equation \ref{effmass}. The latter relates the effective mass of a
packing, a dynamical measure, to the inverse of the dynamical matrix
$H_{ij}$, which is given by the static packing structure.  The test of
equation \ref{effmass} involves two steps: First we perform a direct
dynamical measure of the effective mass by shaking a packing generated
by computer simulations at a given frequency.  For a given shaking
frequency and amplitude $A=1\mu$m, which is very small compared to the
radii and overlap between particles, we record the equilibrium
positions in the packing and shake the bottom wall for a time long
enough to measure the force against the bottom wall.  According to the
time series of the force, we can compute the
effective mass corresponding to the frequency.  For this calculation we
consider a fixed value of damping parameter $\xi$.  Thus, we measure
the force and acceleration of the cup and extract the effective
mass following equation \ref{meff}.  We hold the amplitude constant and
vary the frequency therefore, the acceleration increases with the
frequency.

The real and imaginary part of the DEM simulation are plotted as a
function of $\omega$ in Fig. \ref{TheorySim}. The calculated effective
mass exhibits dominant resonance features at similar frequencies as
that observed experimentally (Hsu et al., 2009; Valenza et al, 2009; Valenza and Johnson, 2012; Hu et al., 2014).   Next, we calculate
$H^{-1}(\omega)$ by using the static positions of the grains before
the shaking in order to employ equation \ref{effmass}.
Figure ~\ref{TheorySim} shows that the DEM and analytical result
(equation \ref{effmass}) are in agreement.  We think it is important to
point out that the analytical prediction (equation \ref{effmass})
accounts for the complex behavior of dissipative granular
media. Moreover, we demonstrate that the agreement is consistently
good over an order of magnitude in damping parameter.

\begin{center}
\begin{large}
{\bf COMPUTATIONAL METHOD FOR TRAJECTORIES}
\end{large}
\end{center}

Given the good agreement demonstrated in Figure \ref{TheorySim} using the two different ways of computing the effective mass of our system, we
calculate the normal modes of the system to test equations
\ref{eq:scaling_xi_c}, \ref{circle} and \ref{parabola}. By using
the static positions of the grains obtained from the packings, we can
solve equation \ref{eigen} to extract the normal modes by following the
method derived by Meirovitch (1987).  In the time domain,
 equations\ref{h} and \ref{eigen} take the form:
 \begin{equation}
\mathbf{M} \mathbf{\ddot{x}} + \mathbf{B} \mathbf{\dot{x}} +
\mathbf{K} \mathbf{x} = 0.
\label{time}
\end{equation}
If we utilize the state vector $\mathbf{y} = [x_1, \cdot, x_N, \dot{x}_1,
  \cdot, \dot{x}_N]^T$,  equation \ref{time} can be recast as:
 \begin{equation} \mathbf{\dot{y}} = \mathbf{A} \mathbf{y},
\end{equation}
where
 \begin{equation}
\mathbf{A}=\left(
  \begin{array}{cc}
    \mathbf{0} &\mathbf{ I} \\
    -\mathbf{M}^{-1}\mathbf{K} & -\mathbf{M}^{-1}\mathbf{B} \\
  \end{array}
\right). \label{eigtransform}\end{equation}
Here, the normal modes frequencies are
 \begin{equation}
\omega_n=i\lambda_n,
\label{lambda}
\end{equation} where
$\lambda_n$ is the n-th eigenvalue of matrix $\mathbf{A}$.  We use the
 methods developed by Lehoucq and Sorensen (1996) and Sorensen (1992)
 to solve for the eigenvalues of $\mathbf{A}$.

%

\begin{center}
\begin{large}
{\bf EFFECT OF DAMPING}
\end{large}
\end{center}

Since the results of our DEM simulations are consistent with theory
and experiment, we proceed to study the effect of the damping
parameter, $\xi$, on the normal mode frequencies.  In particular we
investigate the trajectories of the normal modes in the complex plane
as we change the damping from undamped to critically damped and
beyond. We first calculate the normal modes of the undamped system.
For any single trajectory, we calculate the starting point at the
undamped frequency $\omega_{n0}$. To obtain the trajectory of this
particular normal mode we set an objective function $y=\min[{|
    \lambda_1(\xi+\Delta\xi)-\lambda_1(\xi) |^2,\cdots, |
    \lambda_n(\xi+\Delta\xi)-\lambda_n(\xi)|^2}$, where $\lambda_i$
are given by  equation \ref{lambda}. Then, we increase $\xi$ by a small
amount to $\xi+\Delta\xi$.  Using the `Eig' function in Matlab, we
calculate $y$ for a given damping. Next we employ the multidimensional
unconstrained nonlinear minimization method (`Fminsearch' function in
Matlab) with a starting point $\omega_{n}(\xi)$ to determine
$\omega_n(\xi+\Delta\xi)$ corresponding to $y=0$. It is important to
note that using the starting point $\omega_n(\xi)$, implies that the
solution will be in the vicinity of $\omega_n(\xi)$.  This is the case
when $\Delta\xi$ is small enough, which allows us to determine the
exact trajectory of the normal mode in the complex plane as $\xi$ is
varied.

We first test the method of  equation \ref{eigtransform} to calculate the
normal modes using a small system of 14 particles, for which the modes
can be easily calculated. We compare the modes obtained from
 equation \ref{eigtransform} with a direct numerical calculation of the
eigenvalues of the dynamical matrix $H_{ij}$, ie, we calculate all the
$\omega_n$ which satisfy $\det(\mathbf{H}) =0$. We use a method
derived from expanding the expression for the determinant of
$\mathbf{H}$. The desired roots of the resulting polynomial in
$\omega$ are equal to the eigenvalues of a simple matrix formed from
the coefficients of that polynomial. The eigenvalues are calculated
with the 'Eig' function in Matlab. A potential problem of this
technique appears for large number of particles.  Thus, we use this
technique to validate the use of  equation \ref{lambda} in a small system
of 14 particles, and later we will use  equation \ref{lambda} to calculate
the normal modes for larger systems of 400 particles.

For instance, consider the results of 14 balls interacting via central
forces. The number of normal modes is $6\times 14$ and therefore the
order of the polynomial to calculate the roots is 84. In our
calculations, the different normal mode frequencies, $\omega_n$, vary
by a factor of 10 from the highest to lowest. If one were to evaluate
the polynomial directly, the term $\omega^{84}$ would vary by a factor
$10^{84}$. This is well beyond the accuracy with which a computer
stores numbers. So there may be a problem with the accuracy of the
results obtained by a direct numerical evaluation of the normal modes
via the eigenvalues of $H_{ij}$ for large numbers of particles.

\begin{figure}
\centering{
  \includegraphics[width=0.9\columnwidth]{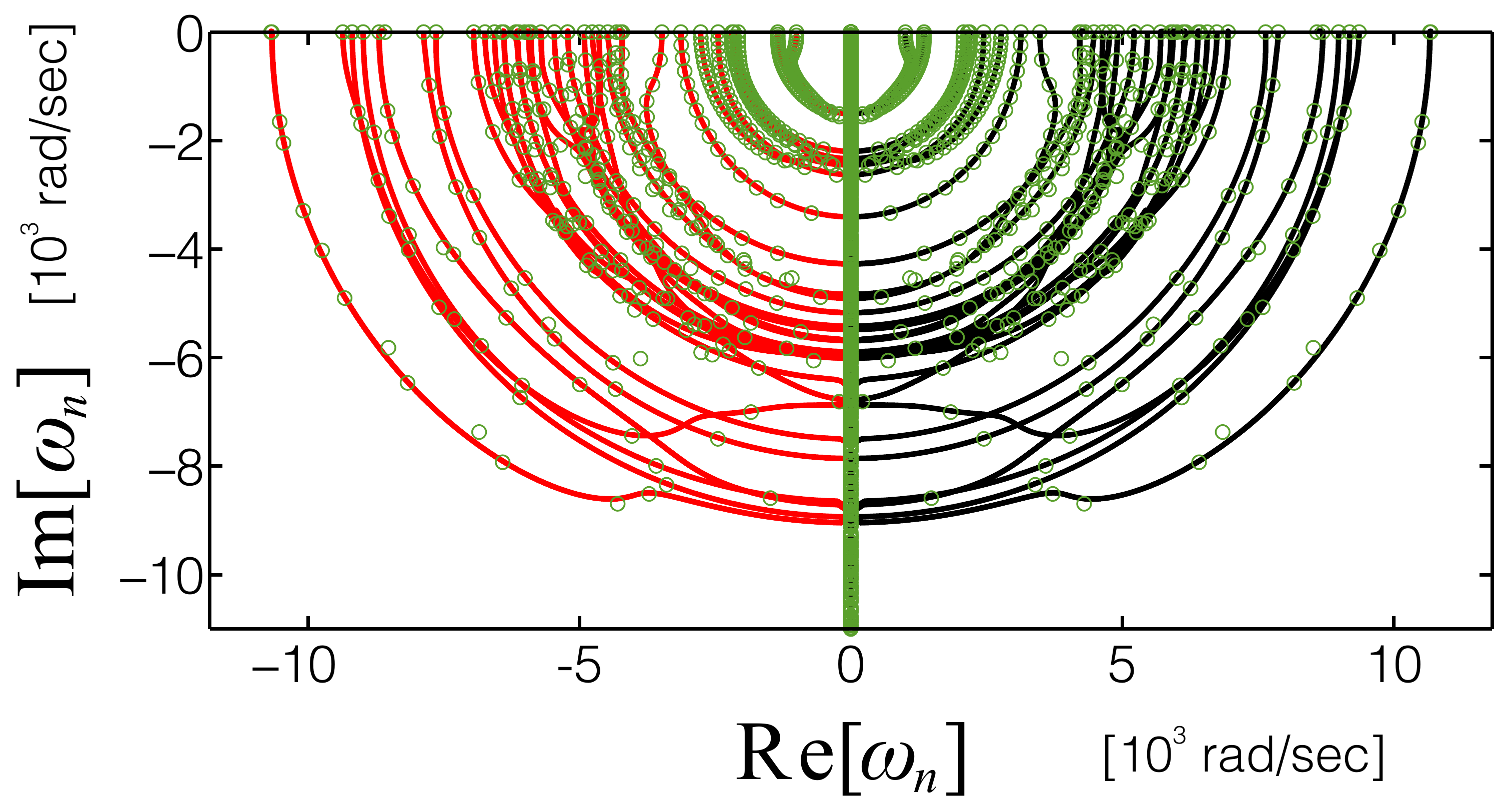}}
\caption{}
\label{14}
\end{figure}

Figure \ref{14} shows a comparison of both methods to calculate the
normal modes as a function of $\xi$ for a system of 14 particles.  The
system is small enough that we can plot the trajectories of all the 84
normal modes. We see that for this system,  equation \ref{lambda} gives
the same result as the direct calculation of normal modes via
eigenvectors of the dynamical matrix $\mathbf{H}$
(equation  \ref{eigen}).  We conclude that the alternative method of
Meirovitch (1987),  equation \ref{lambda}, works well and then we proceed
to use it for large system sizes.

\begin{figure}
\centering{ \includegraphics[width=0.5\columnwidth]{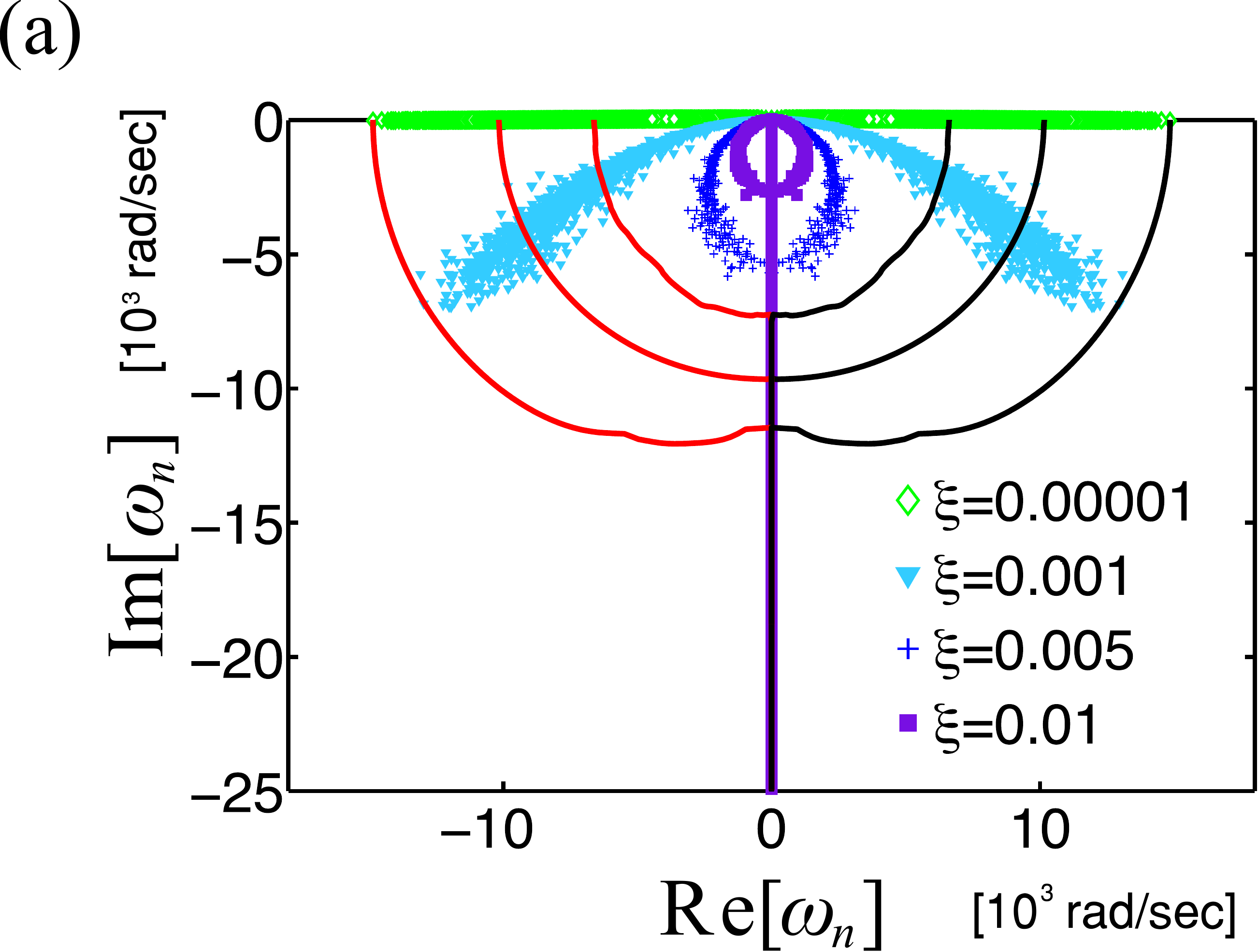}}
\centering{ \includegraphics[width=0.5\columnwidth]{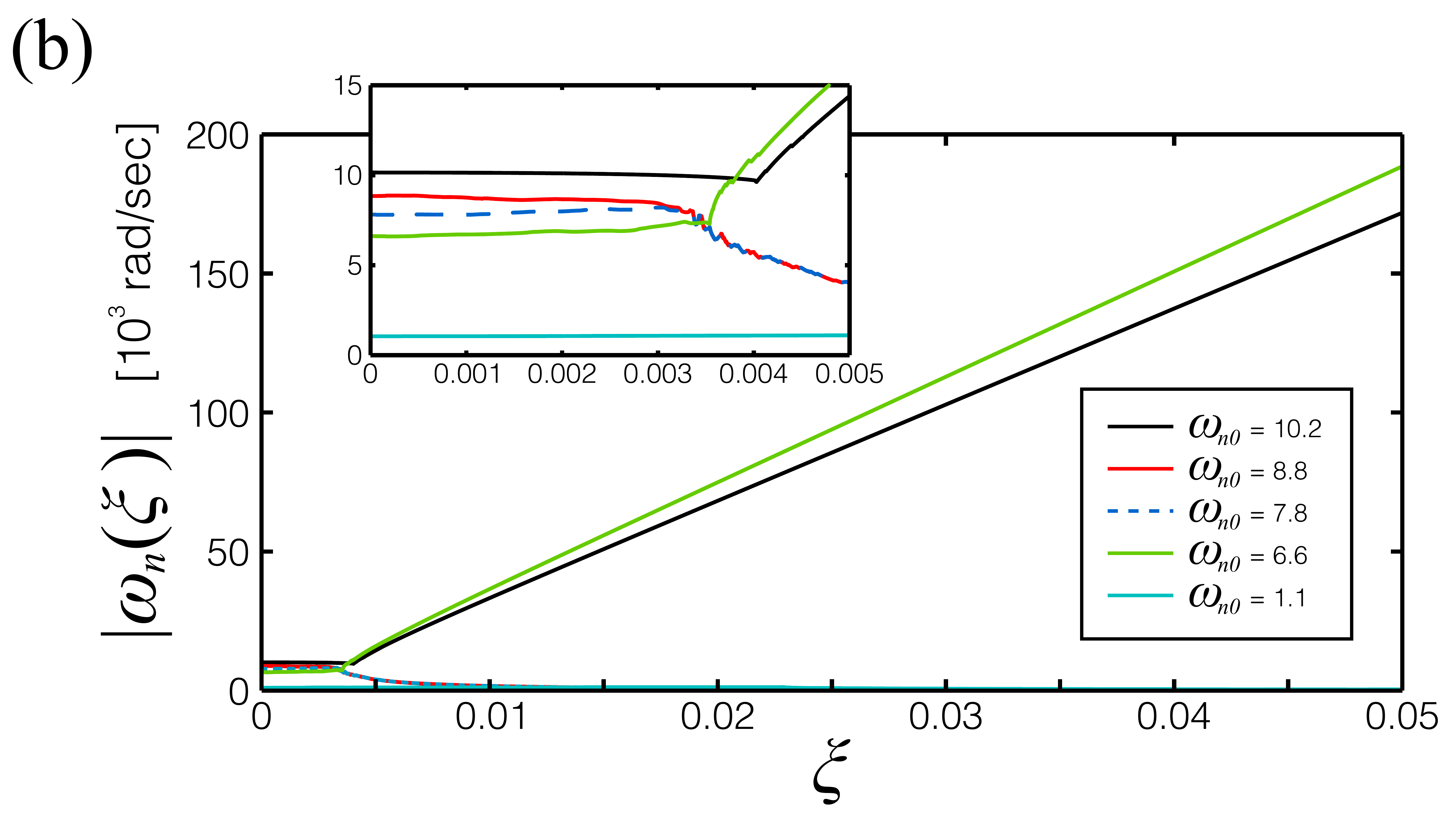}}
\caption{}
\label{fig:menorah}
\end{figure}

Figure ~\ref{fig:menorah}a shows the results of the trajectories of
the normal modes as a function of $\xi$ for a system of $N=400$
particles. The normal modes are calculated via equation
\ref{lambda}. We start with a specific undamped mode $\omega_{n0}$ and
then follow its trajectory as a function of $\xi$ in the
$\Big($Re$[\omega_n(\xi)],$ Im$[\omega_n(\xi)]\Big)$-plane. We follow
the trajectories of specific undamped modes by increasing the damping
parameter $\xi$ in small steps and using the method explained
above. Since the system is large, we do not trace the trajectories of
all 2400 modes. To indicate the general shape of the mode trajectories
we trace that for three modes characterized by undamped frequencies
$\omega_{n0} = 6.5, 10.2$ and 15 $\times \:10^3$ rad/sec, as indicated
by the solid red and black lines in the figure, respectively. In
addition, we also plot all the modes for four values of $\xi$ as
indicated in the figure.

In general, the theoretical predictions equations
\ref{eq:scaling_xi_c}, \ref{circle} and \ref{parabola} are in rather
good agreement with the results in Figure \ref{fig:menorah}; the small
deviations can be traced back to the approximation equation
\ref{propto} which is not exact in the considered packings.  This plot
shows that the modes follow a nearly circular trajectory as we change
$\xi$ (the Menorah-shape traced in continuous black and red lines) as
predicted by equation \ref{eqn:roots_omegan}. When $\xi\ge\xi_c$ the
modes lie on the imaginary axis in agreement with the prediction of
equation \ref{eq:scaling_xi_c}.  A cross-plot of Re[$\omega_n$]
vs. Im[$\omega_n$], for fixed $\xi$ nearly follows equation
\ref{circle}, for modes which are not overdamped.
The latter prediction is valid for small $\xi$ (light blue curve,
$\xi=0.08$). For greater $\xi$ we find the circular plots predicted by
equation \ref{circle} (see blue and purple curves for $\xi=$0.4 and
0.8, respectively). The centers of the circles lie on the imaginary
axis and they are characterized by a radius $\simeq\xi^{-1}$.  The
strong implication, then, is that all the normal mode frequencies for
real granular media lie on or near a circle of radius $R$ centered on
the point $-iR$.  (In our work, here, $R= 1/\xi$.)

 To further test the prediction of circular trajectories, we plot the
 evolution of the absolute value of selected modes for increasing
 values of $\xi$ (Figure \ref{fig:menorah}b). We find that the
 absolute value of the mode is nearly constant, corresponding to the
 circular trajectory, equation \ref{circular} for
 $\xi<\xi_c(\omega_{n0})$, until the critical value of damping unique
 to the individual modes. When $\xi$ approaches $\xi_c$, the absolute
 value of the normal mode frequency is no longer constant and the mode
 becomes overdamped, with the result that $\mid \omega_n \mid$ either
 increases or decreases along the imaginary axis as a function of
 $\xi$.

\begin{figure}
\centering{
  \includegraphics[width=0.9\columnwidth]{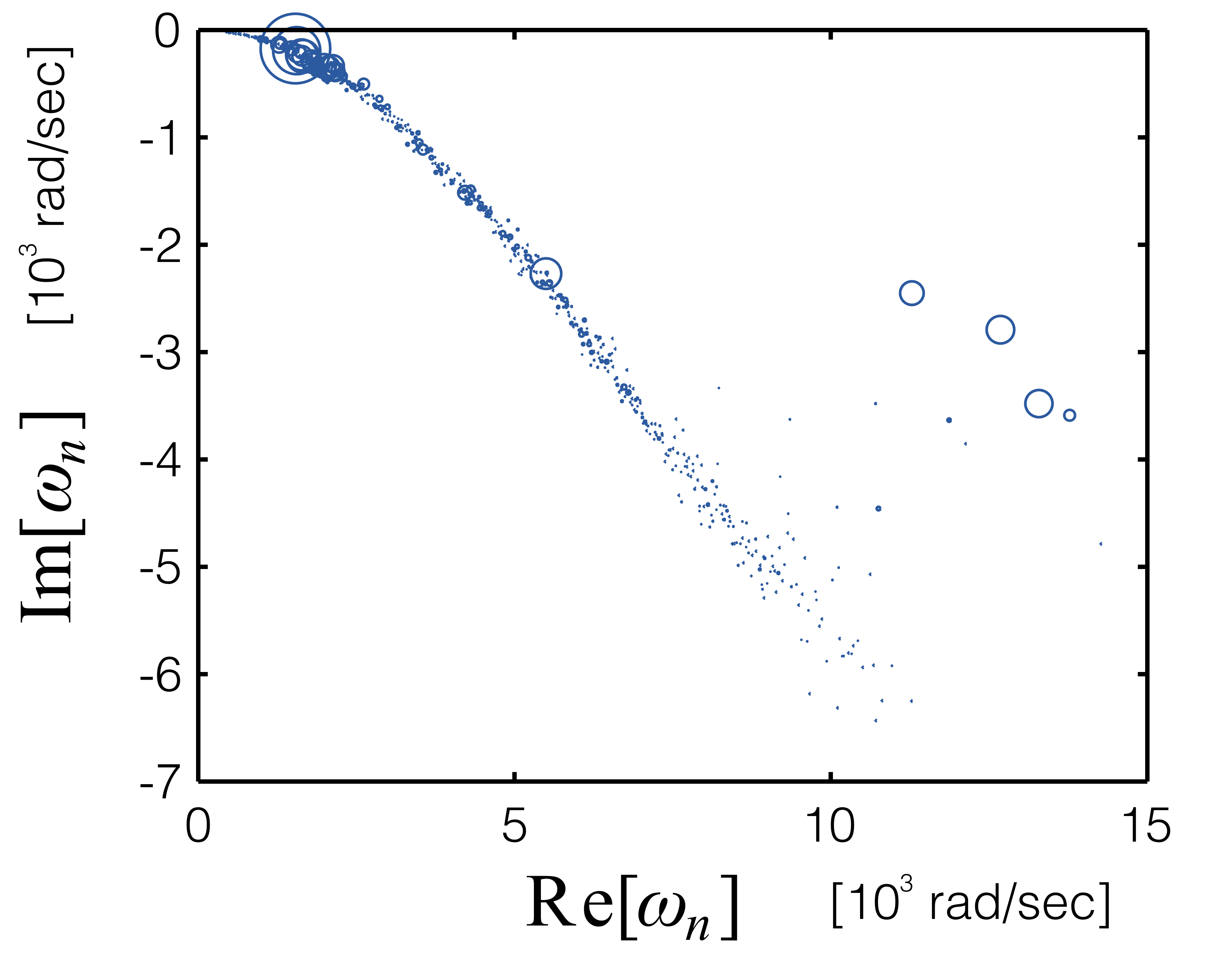}}
\caption{}
\label{parabolic}
\end{figure}

Figure \ref{parabolic} shows the normal modes for a system with a
fixed value of $\xi=0.06$.  In this case, the relative residue of each
mode is indicated by the size of the data point. The residue is
indicative of the contribution of the mode to the effective mass
(Valenza and Johnson, 2012; Hu et al., 2014); large residues
correspond to modes with large contributions to the effective
mass. The modes exhibit a parabolic dependence on ${\rm Re}[\omega_n]$
consistent with equation \ref{parabola}, which is valid for this
system characterized by small $\xi$.  However, we also observe that
several modes with large residue do not lie on the parabola. As
previously noted, the large residue indicates that these outliers make
an important contribution to the effective mass.  They are a direct
manifestation of the breakdown of the approximation equation
\ref{propto}.

\begin{figure}
\centering{
  \includegraphics[width=0.9\columnwidth]{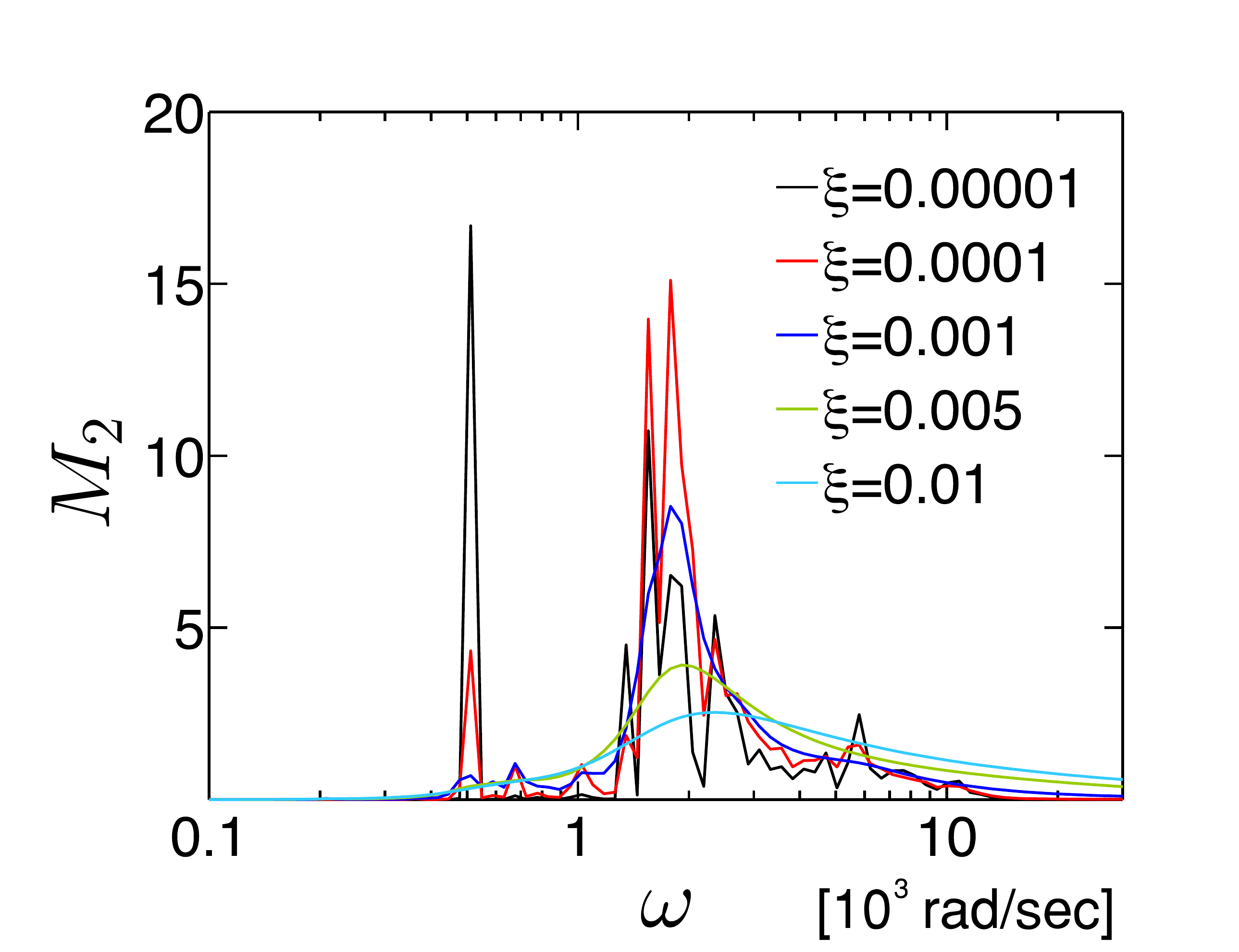}}
\caption{}
\label{imaginarypart}
\end{figure}

Figure ~\ref{imaginarypart} shows the imaginary part of the effective
mass, $M_2$, calculated via DEM, for different values of the damping
parameter $\xi$.  As $\xi$ increases the large amplitude modes spread
out over a larger frequency band, and modes characterized by a small
amplitude, effectively disappear.  That is, for small $\xi$ we find a
large number of peaks in the effective mass. As $\xi$ increases, more
and more peaks disappear. We also find a special low frequency mode
less than 1000 rad/sec with a very large resonance peak indicating
that this frequency contributes a lot to the attenuation of the
medium. When the damping is increased from $\xi=8\times 10^{-4}$ to
$\xi=8 \times 10^{-3}$, the peak almost disappears. Thus, even at
small $\xi$ modes that make important contributions to dissipation can
be overdamped.  The strong attenuation of the first (low-frequency)
peak is observed when increasing the damping parameter, compared to
the high-frequency peaks.  This effect is a consequence of equation
\ref{eq:scaling_xi_c}, which states that the critical damping is
inverse proportional to the frequency of the undamped mode.

\begin{figure}
\centering{
  \includegraphics[width=0.9\columnwidth]{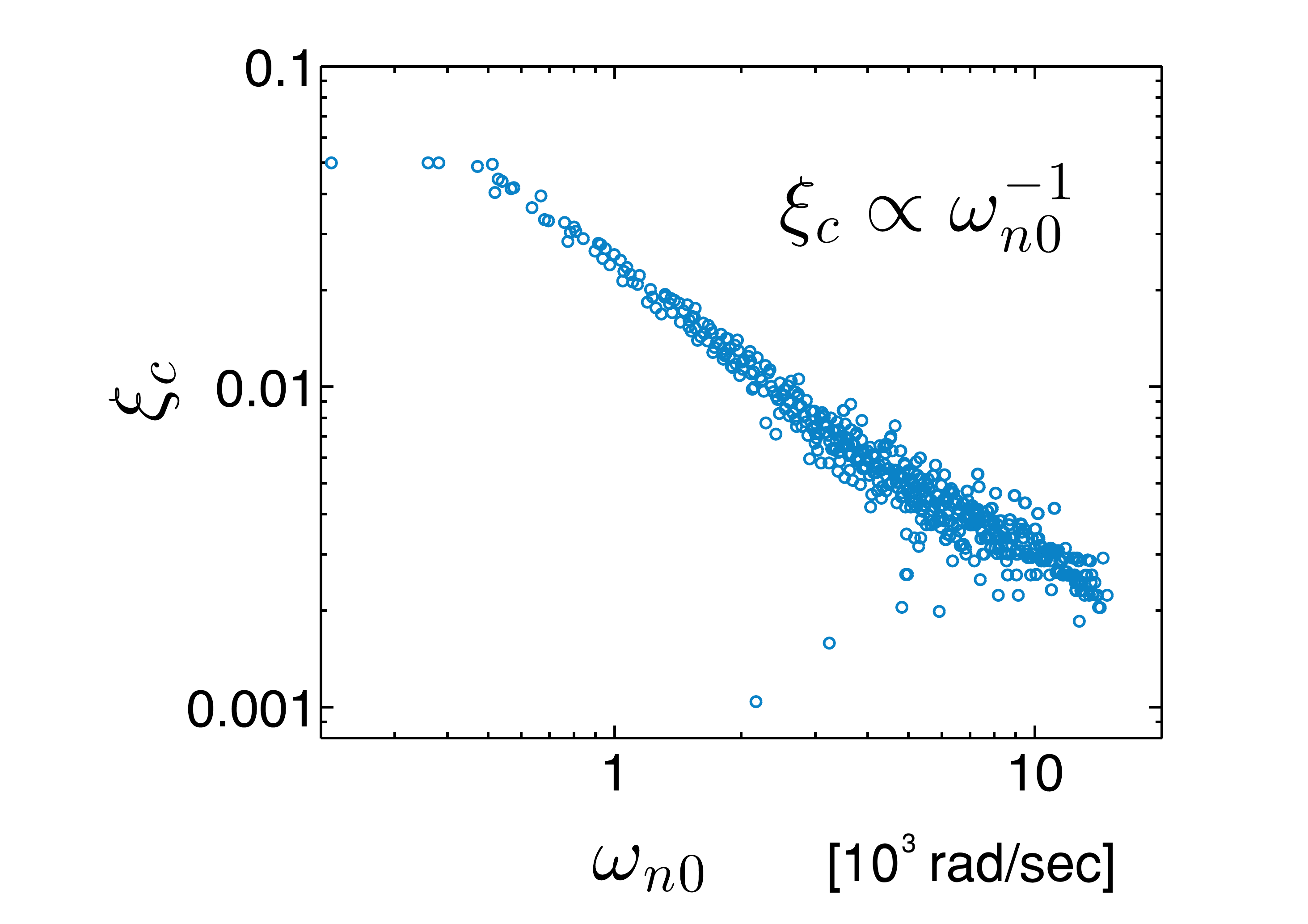}}
\caption{}
\label{fig:omegaXi}
\end{figure}

Finally, we test the scaling of the critical damping $\xi_c$ as a
function of the frequency of the undamped mode $\omega_{n0}$, as
suggested by  equation \ref{eq:scaling_xi_c}.  Figure ~\ref{fig:omegaXi}
shows good agreement between the numerical estimation of
$\xi_c(\omega_{n0})$ and $\omega_{n0}$, confirming the inverse scaling
law of  equation \ref{eq:scaling_xi_c}.  The deviations in the curve in
Fig. \ref{fig:omegaXi} from  equation \ref{eq:scaling_xi_c} are explained
due to the approximation of equation \ref{propto}.  This scaling
law indicates that lower frequency modes are more resistant to
becoming overdamped. Again, some high frequency modes do not obey the
scaling behavior  equation \ref{eq:scaling_xi_c}. These modes have a small
$\xi_c$ value and therefore can be easily dampened.

\begin{center}
\begin{large}
{\bf CONCLUSIONS}
\end{large}
\end{center}

In summary, we have explored the effects of
  interparticle damping on the normal modes in a granular medium.  We
  find good agreement between the DEM simulations and the analytical
  predictions. As the interparticle damping is varied, the normal mode
  trajectories in the complex plane are nearly circular and the
  critical value of the damping parameter, $\xi_c(\omega_n)$, scales
  approximately as $\omega_n^{-1}$.  For a fixed value of the damping
  parameter, $\xi$, the normal mode frequencies lie nearly on the
  circle given by  equation \ref{circle}.  These properties indicate that
  the assumption of proportionality,  equation \ref{propto}, is
  approximately valid, even though it is not strictly true.  The
  strong implication is that these simulation results of ours have
  approximate validity to understanding dissipation effects in real
  granular systems.  Specifically, the indication is that the normal
  mode frequencies of real granular systems lie on or near a circle of
  radius $R$ centered on the point $-iR$, where $R$ is inversely
  proportional to the intergranular damping parameter.

A surprising observation is that there are some very special normal
modes that do not obey this scaling law. The corresponding frequencies
are easy to identify in trajectory plots and are characterized by a
small critical damping parameter.
These outliers can make large contributions to dissipation of acoustic
energy in the granular medium. Overall, the theoretical approach
allows for a systematic investigation of the normal mode frequencies
not only for spherical particles, as presented here, but also for
non-spherical particles as well to understand acoustics and dissipation in all granular
media.

\begin{center}
\begin{large}
{\bf ACKNOWLEDGMENTS}
\end{large}
\end{center}

We acknowledge financial support from DOE Office of Basic Energy
Sciences, Chemical Sciences, Geosciences, and Biosciences Division,
Grant DE-FG02-03ER15458, and NSF-CMMT.  We thank F. Santib\'a\~nez and
S. Reis for discussions and help with the manuscript.

\begin{center}
\begin{large}
{\bf REFERENCES}
\end{large}
\end{center}

\noindent
Alexander, S., 1998,  Amorphous
  solids: their structure, lattice dynamics and
  elasticity: Phys. Rep. {\bf 296}, 65-236.

\bigskip
\noindent
Behringer, R. P. and J. T. Jenkins,  1997, Powders \& Grains, Balkema.

\bigskip
\noindent
Bhaskar, A. 1997, Criticality of damping in multi-degree-of-freedom systems:  J. Appl. Mech. {\bf 64},  387--393.

\bigskip
\noindent
Baule, A., R. Mari, L. Bo. L. Portal, H. A. Makse, 2013, Mean-field
theory for random close packings of axisymmetric particles: Nature
Commun. {\bf 4}, 2194.

\bigskip
\noindent
Baule, A. and H. A. Makse, 2014, Fundamental challenges in packing
problems: from spherical to non-spherical particles: Soft Matter {\bf
  10}, 4423.

\bigskip
\noindent
Bruji\'c, J., P. Wang, C. Song, D. L.  Johnson, O. Sindt, and
H. A. Makse,, 2005, Granular dynamics in compaction and stress
relaxation: Phys. Rev. Lett. {\bf 95}, 128001.

\bigskip
\noindent
 Bruji\'c, J., C. Song, P. Wang, C. Briscoe, G. Marty, and
 H. A. Makse, 2007, Fluorescent contacts measure the coordination
 number and entropy of a 3D jammed emulsion packing:
 Phys. Rev. Lett. {\bf 98}, 248001.

\bigskip
\noindent
 Brunet, Th., X. Jia,  and P. Mills, 2008, Mechanisms for acoustic absorption in dry and weakly wet granular media: Phys. Rev. Lett. {\bf 101}, 138001.

\bigskip
\noindent
 Caughey, T. K., 1960, Classical normal modes in damped linear dynamic systems:  J. Appl. Mech. {\bf 27}, 269-271; Caughey, T. K.  and
  M. E. J. O'Kelly, 1965, Classical normal modes in damped linear dynamic systems: J. Appl. Mech. {\bf 32}, 583-588

\bigskip
\noindent
  Chen, Y.-C., I. Ishibashi, and J. T. Jenkins, 1988,  Dynamic shear modulus and fabric: part I, depositional and induced anisotropy: {\it G\'eotechnique} {\bf 38}, 25-32.

\bigskip
\noindent
Cundall, P. A. and O. D. L. Strack, 1979,  A Discrete Numerical Model for Granular
  Assemblies: G\'eotechnique {\bf 29}, 47-65.

\bigskip
\noindent
 de Gennes, P.-G.,  1996, Static
  compression of a granular medium: The ``Soft Shell" model:   Europhys. Lett. {\bf 35}, 145-149.

\bigskip
\noindent
Digby, P. J., 1981, The effective
  elastic moduli of porous granular rocks:  J.  Appl. Mech.
  {\bf 48}, 803-808.

\bigskip
\noindent
 Domenico, S. N., 1977,
 Elastic properties of unconsolidated porous sand reservoirs:  Geophysics {\bf 42}, 1339-1368.

\bigskip
\noindent
 Garvey, S. D.,
  J. E. T. Penny, and M. I. Friswell, 1998, The relationship between the real and the imaginary parts of complex modes:  J. Sound \& Vib. {\bf 212}, 75-83.

\bigskip
\noindent
Goddard, J. D., 1990, Nonlinear
  Elasticity and Pressure-dependent Wave Speeds in Granular Materials:
  Proc. R. Soc. Lond. A {\bf 430}, 105-131.

\bigskip
\noindent
Guyer, R. A. and
  P. A. Johnson, 1999, Nonlinear mesoscopic elasticity: Evidence for a new
  class of materials: Phys. Today {\bf 52}, 30.

\bigskip
\noindent
Henann, D. L., J. J. Valenza, D. L. Johnson, and K. Kamrin, 2013,  Small amplitude acoustics in bulk granular media: Phys. Rev. E {\bf 88} 042205.

 \bigskip
\noindent
Hsu, C-J., D. L. Johnson, R. A. Ingale, J. J. Valenza, N. Gland, and H. A. Makse, 2009, Dynamic effective mass of granular
  media: Phys. Rev. Lett. {\bf 102}, 058001.

\bigskip
\noindent
Hu, Y., D. L. Johnson, J. J. Valenza, F. Santib\'a\~nez, and
H. A. Makse, 2014, Stress-dependent normal mode frequencies from the
effective mass of granular matter: Phys. Rev. E {\bf 89}, 062202.

\bigskip
\noindent
Inman, D. J. and A. N.  Andry, 1980, Some results on the nature of
eigenvalues of discrete damped systems: J. Appl. Mech. {\bf 47},
927--930.

\bigskip
\noindent
Jin, Y. and H. A. Makse, 2010, A first-order phase transition defines
the random close packing of hard spheres: Physica A {\bf 389},
5362-5379 (2010).

\bigskip
\noindent
Johnson, D. L., L. Schwartz, D. Elata, J. G. Berryman, B. Hornby,
A. N. Norris, 1998, Linear and nonlinear elasticity of granular media:
Stress-Induced anisotropy of a random sphere pack: ASME
J. Appl. Mech. {\bf 65}, 380-388.

\bigskip
\noindent
Johnson, K. L., 1985,   Contact Mechanics: Cambridge University Press.

\bigskip
\noindent
Johnson, P. A.,
  B. Zinszner, and P. N. J. Rasolofosaon, 1996,  Resonance and nonlinear
  elastic phenomena in rock: J. Geophys. Res. {\bf 101}, 11553-11564.

\bigskip
\noindent
Landau, L. D. and E. M. Lifshitz, 1970,  Theory of Elasticity: Pergamon.

\bigskip
\noindent
Lehoucq, R. B. and D. C. Sorensen, 1996, Deflation techniques for an
implicitly re-started Arnoldi iteration: SIAM J. Matrix Analysis and
Applications {\bf 17}, 789-821.

\bigskip
\noindent
Magnanimo, V., L. La Ragione, J. T. Jenkins, P. Wang, and H. A. Makse,
2008, Characterizing the shear and bulk moduli of an idealized
granular material: Europhys. Lett. {\bf 81}, 34006.

\bigskip
\noindent
Makse, H. A., N. Gland, D. L.  Johnson, and L.M. Schwartz, 1999, Why
effective medium fails in granular materials:
Phys. Rev. Lett. \textbf{83}, 5070-5073.

\bigskip
\noindent
Makse, H. A., D. L. Johnson, and L. M.  Schwartz, 2000, Packing of compressible granular materials: Phys. Rev.  Lett. {\bf 84}, 4160-4163.

\bigskip
\noindent
Makse, H. A., J. Bruji\'c, and S. F. Edwards, 2004, Statistical
mechanics of jammed matter: in The Physics of Granular Media,
H. Hinrichsen and D. E. Wolf (eds.)  Wiley.

\bigskip
\noindent
Meirovitch, L. , 1987, Principles and techniques of vibrations:
Prentice-Hall.

\bigskip
\noindent
Norris, A. N. and
  D. L. Johnson, 1997,  Nonlinear elasticity of granular media: J. Appl. Mech. {\bf 64}, 39-49.

\bigskip
\noindent
O'Connell, R. J., and B.  Budiansky, 1977, Viscoelastic properties of
fluid-saturated cracked solids: J. Geophys. Res. B {\bf 82},
5719-5735.

\bigskip
\noindent
O'Hern, C. S., L. E. Silbert, A. J.  Liu, and S. R.  Nagel, 2003,
Jamming at zero temperature and zero applied stress: The epitome of
disorder: Phys. Rev. E {\bf 68}, 011306.

\bigskip
\noindent
Shafer J., S. Dippel,and D. E. Wolf, 1996, Force schemes in
simulations of granular materials: J. Phys. I (France) {\bf 6}, 5-20.

\bigskip
\noindent
Silbert, L. E., A. J. Liu, and S. R.  Nagel, 2005, Vibrations and
diverging length scales near the unjamming transition:
Phys. Rev. Lett. {\bf 95}, 098301.

\bigskip
\noindent
Somfai, E., {\it et al.}, 2007, Critical and noncritical jamming of
frictional grains: Phys. Rev. E {\bf 75}, 020301R.

\bigskip
\noindent
Song, C., P. Wang, and H. A. Makse, 2008, Phase diagram for jammed
matter: Nature {\bf 453}, 629-632.

\bigskip
\noindent
Sorensen, D. C., 1992, Implicit Application of polynomial filters in a
k-step Arnoldi method: SIAM J. Matrix Analysis and Applications, {\bf
  13}, 357-385.

\bigskip
\noindent
Sun, J. C., H. B. Sun, L. C. Chow, and E. J. Richards, 1986,
Predictions of total loss factors of structures, part II: Loss factors
of sand-filled structure: J. Sound and Vib. {\bf 104}, 243-257.

\bigskip
\noindent
Valenza, J. J., C.-J. Hsu, R. Ingale, N. Gland, H. A. Makse, and
D. L. Johnson, 2009, Dynamic effective mass of granular media and the
attenuation of structure-borne sound: Phys. Rev. E {\bf 80}, 051304.

 \bigskip
\noindent
Valenza, J. J. and D. L.  Johnson, 2012, Normal-mode spectrum of
finite-sized granular systems: The effects of fluid viscosity at the
grain contacts: Phys. Rev. E {\bf 85}, 041302.

 \bigskip
\noindent
Walton, K., 1987, The effective elastic moduli of a random packing of
spheres: J. Mech. Phys. Solids {\bf 35}, 213-226.

\bigskip
\noindent
Winkler, K. W., 1983, Contact stiffness in granular porous materials:
Comparison between theory and experiments: Geophys. Res. Lett.  {\bf
  10}, 1073-1076.

\bigskip
\noindent
Wyart, M., S. R. Nagel, and T. A.  Witten, 2005a, Geometric origin of
excess low-frequency vibrational modes in weakly connected amorphous
solids: Europhys. Lett. {\bf 72}, 486-492.

\bigskip
\noindent
Wyart, M., L. E.  Silbert, S. R. Nagel S. R. and T. A.  Witten, 2005b,
Effects of compression on the vibrational modes of marginally jammed
solids: Phys. Rev. E {\bf 72}, 051306.

\clearpage

FIG. \ref{fig:packing}. Schematic view of the simulation box and
particles. The simulations consider periodic boundary conditions. The
walls are made of fixed particles of the same features of the
particles in the bulk. The figure shows a system prepared after the
relaxation under gravity has finished. The walls are shaken during the
dynamical calculation of the effective mass with an acceleration
$a(\omega)$ and the force $F(\omega)$ is measured at the bottom of the
cup to obtain the effective mass via  equation \ref{meff}.

FIG. \ref{TheorySim}.Comparison of theoretical estimation of the
effective mass via  equation \ref{effmass} and the direct dynamical
measurements using the numerically generated packings with DEM by
shaking the packing at a given frequency $\omega$.  DEM results are
shown by the symbols for two damping parameters $\xi=0.08$ms and
$\xi=0.8$ms and separately for the real and imaginary part of the
effective mass $\tilde{M}(\omega)$, as indicated. The DEM shaking
simulation results are obtained from the frequency sweep with
amplitude $A=1 \mu$m. The theory lines correspond to the solution of
the right hand side of  equation \ref{effmass}: $m [H^{-1}(\omega)]_{ij}
K_{j\omega}$, where $H^{-1}$ is calculated as discussed in the
text. The two different methods of computation are in good agreement with each other.

FIG. \ref{14}. Trajectories of all the 84 resonance frequencies for a
small packing of 14 particles. This plot shows the comparison between
the trajectories of $\omega_n(\xi)$ calculated by the Meirovitch
method of  equation \ref{eigtransform} (circles) with the direct
determination of the determinant of the dynamical matrix $\mathbf{H}$
(line trajectories). We confirm that both methods gives the same
result for the normal frequencies. This result indicates that
 equation \ref{eigtransform} is an efficient way to calculate the normal
modes, and it is used then in the calculations for larger system sizes
in the rest of the paper. The trajectories follow the Menorah-like
shapes which are further investigated in Fig. \ref{fig:menorah} for a
larger system.

FIG. \ref{fig:menorah}. (a) Locus of all the complex-valued normal
mode frequencies of our large system for four different values of the
damping parameter, $\xi$.  The normal modes are calculated via the
solution of equation \ref{eigtransform}.  Also shown are the
trajectories of three of the normal mode frequencies as $\xi$ is
varied from low to high values.  The frequencies approximately follow
equations \ref{circular} and \ref{circle}.  The stated values for
$\xi$ are in millisec.  (b)$\mid \omega_n(\xi)\mid $ vs. $\xi$ shown
for a few selected modes obtained from (a) which shows how nearly
circular the trajectories are up to the point of critical damping.

FIG. \ref{parabolic}. Distribution of normal mode frequencies in the
complex plane for a given fixed small damping $\xi=0.06$ms for
$N=400$. The size of the symbol is proportional to the resonance mode
residue.


FIG. \ref{imaginarypart}. Imaginary effective mass calculated via the
direct dynamical shaking of the medium for increasing values of $\xi$,
in millisec, for $N=400$.


FIG. \ref{fig:omegaXi}. Critical damping $\xi_c$ as function of the
normal mode frequency $\omega_{n0}$ of the undamped mode for
$N=400$. The theoretical prediction $\xi_c\propto \omega_{n0}^{-1}$ in
equation \ref{eq:scaling_xi_c} is in satisfactory statistical
agreement with the numerical results.

\clearpage

\clearpage

\clearpage

\clearpage

\clearpage

\clearpage

\clearpage

\end{document}